\documentstyle[prl,aps,twocolumn,epsf]{revtex}
\def\break#1{\pagebreak \vspace*{#1}}
\begin{document}

\draft

\title{Self-organized digital disorder of Davydov's beta kink}

\author{Haret C. Rosu
}

\address{
{Instituto de F\'{\i}sica de la Universidad de Guanajuato, Apdo Postal
E-143, Le\'on, Guanajuato, M\'exico}\\
}

\maketitle
\widetext

\begin{abstract}

I discuss the digital disorder introduced by Rosu and
Canessa [Phys. Rev. E {\bf 47}, R3818 (1993)] in the Davydov model of energy
diffusion along $\alpha$-helix protein chains.
The digitally disordered Davydov beta kinks display self-organized
features, i.e.,
power law correlations both in time and space, that may be attributed to
incipient dynamic structural changes of the protein chain as a consequence
of coarse-graining the fluctuations due to microscopic degrees of freedom.
In this paper, I
provide a simple semiconductor model for the flicker noise and also comment
on the multifractality
that one may associate with the protein chains by means of digital disorder.

\end{abstract}

\pacs{PACS numbers: 87.15.-v, 05.40.+j, 71.38.+i\hfill
LAA number: cond-mat/9410034}

\narrowtext

\section{Introduction}


In a previous paper \cite{RC93}, 
Canessa and the present author, stimulated by the
findings of Voss concerning 1/f noise in long samples of DNA sequences
from GenBank \cite{V92}, obtained that type of noise in the Davydov's model
\cite{Dav85}
through a simple digital dynamics of $\alpha$-helix chains.
I recall that digital dynamics/disorder, i.e.,
dynamics in digital time can also lead to other types of complex behavior,
as for example to broken symmetries, oscillations, and chaos \cite{Ch83}.
The aim of the paper 
is to present possible interpretations of the chain dynamics as affected
by the simple digital dynamics of \cite{RC93}
and discuss in a heuristic manner the emerging physical picture.

The outline of the paper is as follows. After reviewing the
Davydov model and solutions and shortly discussing
the digital noise of the so-called $\beta$-kink in the next section,
I provide in section 3
a simple physical picture for the flicker noise in terms of a partial
site-trapping of the kink. In section 4, I discuss heuristically the kink
multifractality, and I end up with some concluding remarks.

\section{ Davydov Model}

The $\alpha$-helix is the most
common secondary structure of proteins entailing three spines
 in the longitudinal z-direction that we consider infinite in
extent and having the peptide sequence $(---H-N-C=O)_{n}$, $n=\infty$,
where the dashed lines represent hydrogen bonds. There are side
radicals and their order is characteristic to each protein. The
3 spines are weaved together in a sort of incommensurate quasi-one
\break{1.87in}
dimensional structure, but here we shall consider only the simple
one-spine chain, containing the peptide groups as molecular units.

In the Davydov model the main assumption is that a
substantial part
of the chemical energy ( $\epsilon =0.422\;eV$) released in the
hydrolysis of adenosine
triphosphate (ATP) turns into vibrational energy
($\epsilon _{0}=0.205\;eV$) of the self-trapped amide-I
(C=O stretching) mode of the peptide unit. The amplitude of the
amide mode is self-trapped in the form of a sech- envelope soliton
whenever there is a balance between
the dipolar nearest-neighbour interaction and an admittedly
quite strong nonlinear amide-phonon
interaction. The energy transfer in the standard Davydov model is through
the nonradiative resonant dipole- dipole interaction and a rather
ambiguous vibrational-acoustic coherent state, the so-called $D_{1}$
{\em ansatz}.
It is a mixture of quantum and classical Hamiltonian methods that has
been deeply scrutinized in the literature \cite{Blw86}. Davydov
Hamiltonian can be written in the form
\begin{eqnarray}
    H_{D} & = & H_{C=O}+H_{ph}+H_{int}\; \; \label{eq:d1}\\
\;\;H_{C=O} & = & \sum_{n} {\epsilon _{0}B_{n}^{\dagger}B_{n}-
J(B_{n}^{\dagger}B_{n+1}+ B_{n+1}^{\dagger}B_{n})} \;\; \label{eq:d2}\\
   H_{ph} & = & \sum_{q}\hbar\Omega_{q}(b_{q}^{\dagger}b_{q}+\frac{1}{2})
\; \; \label{eq:d3}\\
H_{int} & = & \frac{1}{\sqrt{N}}
\sum_{q,n}\chi(q)e^{iqnR}B_{n}^{\dagger}B_{n}(b_{q}+b_{-q}^{\dagger})
\; \; \label{eq:d4}
\end{eqnarray}
The capital and small operators are vibrational and phonon ones, respectively.
Brown \cite{Br88} has shown in a clear way that Davydov Hamiltonian is
a particular case of the general Fr\"ohlich Hamiltonian of polaronic
systems \cite{Fro54}
\begin{equation}\label{eq:r3}
H_{F}=\sum _{mn}J_{mn}a_{m}^{\dagger}a_{n}+
\sum _{q} \hbar \omega _{q}b_{q}^{\dagger}b_{q}+
\sum _{qn} \hbar \omega _{q}(\chi _{n}^{q}b_{q}^{\dagger}
+\chi_{n}^{q*}b_{q})a_{n}^{\dagger}a_{n}
\end{equation}
with $D_{1}$ states satisfying the Schr\"odinger equation of the
Fr\"ohlich Hamiltonian in the limit $J_{mn}=0$ and another $D_{2}$
{\em ansatz} valid for Schr\"odinger evolution in the limit
$\chi _{r}^{q}=0$.

The continuous limit, nonlinear Schr\"odinger (NLS) subsonic soliton
solutions of the energy transport coming out from the Davydov model are
\begin{eqnarray}
\alpha(\xi) & = & \sqrt{\mu /2}
e^{[\frac{i}{\hbar}[\frac{\hbar ^{2}v_{s}x}{2JR^{2}}-E_{s}t]]}
\cosh ^{-1}(\frac{\mu}{R}\xi) \; \; \label{eq:dd1} \\
\rho (\xi) & = & \frac{\chi \gamma _{s}^{2}}{w}
sech ^{-2}(\frac{\mu}{R}\xi) \; \; \label{eq:dd2} \\
\beta (\xi) & = & \frac{\chi \gamma _{s}^{2}}{w}(1-\tanh(Q\xi))
\; \; \label{eq:dd3}
\end{eqnarray}
where $\xi =x-v_{s}t$ is the moving frame coordinate, $J$ is
the hopping (dipole-dipole) constant, $\chi$ is the nonlinear
dipole-phonon coupling parameter, $\gamma _{s}=1/\sqrt{1-s^{2}}$
($s=v_{s}/v_{a}$) is the soliton
`relativistic' factor, $w$ is the elasticity constant of the chain,
$\mu =\chi ^{2} \gamma _{s}^{2}/Jw$,
$Q=MR\chi ^{2}\gamma _{s}^{2}/2w\hbar ^{2}$, $E_{s}=\epsilon _{0}-2J
+\hbar ^{2} v_{s}^{2}/4JR^{2}- J\mu ^{2}/3$, and
$\epsilon _{0}\approx 0.205 eV$, the quantum energy of the amide
dipole oscillator.
The first soliton is the vibrational soliton in which one may
remark the soliton energy $E_{s}$ self-trapped by the carrier wave.
The second solution is related to the local deformation produced by the
vibrational soliton in the lattice.
The $\beta$-kink is a domain-wall configuration of the displacements
of the peptide groups from their equilibrium positions. It has an
enhanced stability of topological origin since all the peptide groups
from the right side of the kink ($\xi > 0$ ) are in nondisplaced
positions
whereas all the peptide groups at the left ($\xi < 0$) are displaced
by the same amount $\beta _{0}=\frac{2\chi \gamma _{s}^{2}}{w}$. In
order to destroy the domain wall configuration, one should first turn the
left peptide groups to their initial position. We think of
the Davydov $\beta$-kink as an interphase boundary for the
non-equilibrium transition from the Davydov dynamical regime of the
polypeptide chain to a dynamic `ferroelectric' phase of the chain.
In the literature on ferroelectricity it is common to consider the
interfacial boundary as the kink solution of a time-dependent Ginzburg-Landau
(GL) equation \cite{Gor83}. Our interpretation is based on the fact that
for low subsonic regime ($s^{2}\ll 1$) the Davydov kink is just the
complement of the GL kink (i.e., $K_{D}\propto (1-K_{GL})$). Since we
are in a non-equilibrium situation the more precise terminology for these
kinks is dynamic
interphase boundaries or interfacial patterns. One would like to
study their morphology during the growth. This is a difficult task
since we are in a more complicated case as compared to the simple
solid on solid model, or the kinetic Ising model, where the
width of the interface is given in terms of the nearest neighbor
exchange interaction \cite{Cho88}, corresponding to the $J$ parameter
in the Davydov model. The width of the $\beta$-kink is determined
by two parameters, namely the $J$ of nearest neighbors and the $\chi$
parameter of
the nonlinear interaction, controling the interfacial morphology.

Perhaps, one should notice the formal analogy between the form of the
Davydov kink and the Glauber transition rate in one-dimensional
spin chains. The message of this analogy is that the kink is just
a step structural function which is required by a Hamiltonian
evolution and by a detailed balance condition in the spatial coordinate.

The digital disorder introduced in \cite{RC93} is due to small random
displacements of the instantaneous kink position, which {\em a priori}
may appear to be
equivalent to small fluctuations of the kink velocity. The procedure
is as follows. The centre of mass of the kink in the moving frame
is taken as the spatial origin and small random excursions $\Delta x$ are
allowed between $\pm 1$. The time correlations of such fluctuations
are estimated by means of the dimensionless noise power spectrum, which
usually characterize random processes of the quantity chosen as
probability density, in our case the $\beta$- kink function. Thus the
temporal correlations at the scale $0\;<t\;<\tau\approx 1/f$ are
measured by \cite{V92}
\begin{equation}\label{eq:r2}
S_{\beta}(f)\propto \frac{1}{\tau}
|\int _{0}^{\tau}\beta (\xi)e^{2\pi ift}dt|^{2}
\end{equation}
When calculated with a standard fast
Fourier transform (FFT) algorithm, $S_{\beta}(f)$ displayed clearly 1/f
noise \cite{RC93}.
Moreover, the fact that the hyperbolic tangent is the solution of the
GL equation with real coefficients leads to
power law correlations in the space domain through a multifractal formalism.
But power law correlations both in time and space are necessary requisites
of self-organized criticality (SOC) \cite{soc87}. The digital noise
has its origin in the removed microscopic degrees
of freedom, for instance the type of lattice, when one is passing to the
continuous NLS limit. In this sense, the digital noise is an internal one
in the terminology of the SOC literature \cite{di}, while the disorders coming
from varying the parameters in the Davydov model, which were studied
some time ago by F\"orner \cite{f91},
belong to the class of external noises. D\'{\i}az-Guilera \cite{di} showed
that the two types of noises can be considered as different limits of an
Ornstein-Uhlenbeck stochastic process.
Actually, the digital noise is very close to a proposal
of Cruzeiro-Hansson \cite{26}, who has argued that the
thermal destabilization
of the Davydov soliton at low temperatures starts through a cascade of
disordered states in the displacements, of intermediate energy between the
soliton and the exciton states. At higher (biological) temperatures, the
disorder in the
displacements is transmitted to the vibrational soliton which in this way
is destroyed, although the states populated in the decay preserve the
localization of the amide I excitation \cite{27}. As a matter of fact,
experimental evidence provided by optical calorimetry shows that phonons
and vibrations heat up at different rates \cite{28}.

\section{Model for the flicker noise of the $\beta$-kink}

A simple interpretation of 1/f noise may be given in terms of a partial
site-trapping of the $\beta$-kink. This
trapping modulates the mobility of the domain wall and is correlated
with the carrier- wave self-trapping. The toy model is adapted from
semiconductor physics \cite{Z59}. We shall consider the
$\alpha$-chain to be
at a slightly non-zero temperature, say $T\leq 11K$
\cite{29}, where the Davydov system is still not washed out by thermal effects.
Let us suppose a sort of
site- trapping changing the kink mobility by rectangular pulses,
whose duration $\tau$ is the trapping time. Furthermore, all pulses
can be considered statistically independent and thus their distribution
is Poisson $p(t)=r\exp (-r\cdot t)$,
where $t$ is the time interval between succesive pulses and $r$ is the
average number of pulses per unit time. Next we shall consider the
distribution function of the trapping times $\tau$, due to a single
site-trap to be of the form
$g(\tau) =\frac{1}{\tau _{0}} \exp(-\tau/\tau _{0})$,
where $\tau _{0}$ is the average trapping time. The last assumption is
that different site-traps have different $\tau _{0}$'s, which are
related to the trapping energy E in the thermal way,
$\tau _{0} =K \exp (-E/kT)$,
where K is a constant, and E is distributed among the trap ensemble
according to a square distribution spectrum between two limiting
energies $E_{1}$ and $E_{2}$. From the thermal assumption one obtains
\begin{equation}\label{eq:r10}
P(\tau _{0})=\frac{1}{\log(\tau _{2}/\tau _{1})}\cdot\frac{1}{\tau _{0}}
\end{equation}
valid for $\tau _{1}\;< \tau _{0}\;< \tau _{2}$, where
$\tau _{1}=K\exp(-E_{1}/kT)$ and $\tau _{2}=\exp(-E_{2}/kT)$.
The power spectrum of a time series of rectangular pulses with the
same duration $\tau$ and Poisson distributed, was obtained long ago
by Gisolf \cite{Gi49}
\begin{equation}\label{eq:r11}
S^{g}_{\omega}=\;
const\cdot\tau\Big(\frac{\sin(\omega \tau /2)}{\omega \tau /2}\Big)^{2}
\end{equation}
By combining $g(\tau)$ with Eq.(\ref{eq:r11}),
Van der Ziel \cite{Z59} obtained a shot-noise power spectrum
\begin{equation}\label{eq:r12}
S^{s}_{\omega}=const\cdot\frac{\tau _{0}}{1+\omega ^{2}\tau _{0}^{2}}
\end{equation}
One gets the 1/f noise by averaging the shot spectrum
with the square distribution of the trapping times Eq.(\ref{eq:r10})
\begin{equation}\label{eq:r13}
S_{flicker}=
\frac{kT}{E_{2}-E_{1}}\frac{1}{\omega}\Bigg[\arctan (\omega \tau _{2})
-\;\arctan (\omega \tau _{1})\Bigg]
\end{equation}
Of course, the simple flicker mechanism we have used here was more
for illustrative purposes, and further work is required to clarify
the nature of such noise in protein chains. For example, the flicker noise
can be explained alternatively by means of a scaled Langevin equation as
introduced by Kogama and Hara \cite{31}. For that, it is sufficient to consider
the same scaling parameter both in the Langevin variable and in the attached
noise as was shown in their paper.

\section{Multifractality of $\beta$- kinks}

The implicit occurence of the GL kink in the Davydov
$\beta$ displacements is essential for discussing the multifractality issue.
Previously, Brax \cite{Br92} showed
the equivalence of the GL equation with real coefficients
under random initial conditions and the linear heat equation with
Gaussian random potential and made a multifractal analysis of the
problem of direct relevance for our study. As is well-known the GL equation,
which is a cubic
reaction- difussion equation, describes phenomenologically the
evolution of the order parameter in superconductive phase transitions,
and models also spatial and time fluctuations of systems near
Hopf bifurcations.
In 1972, Scalapino, Sears and Ferrell \cite{Sca72} studied in
detail the statistical mechanics of one-dimensional GL
fields. They remarked that such fields can describe the dynamical
behaviour of nearly-ordered systems which are not undergoing sharp phase
transitions, and conjectured that the real-field case may have
application in some organic chain systems. Our approach is just
a further confirmation of that idea. As a matter of fact,
in the D'Alambert variable ($\xi=x-v_{s}t$), the GL kink $K_{GL}$ is
the solution of a GL equation of the type
\begin{equation}\label{eq:r20}
\frac{\partial K_{GL}}{\partial t}= \partial ^{2} K_{GL}/\partial x^{2}
+p_1 K_{GL}-p_3 K_{GL}^{3}
\end{equation}
with real $p_1$ and $p_3$ coefficients.
Such GL equations are typical for the structural
phase-transitions \cite{Gor83} in equilibrium situations, but
they can be used also in nonequilibrium/driven systems. Moreover, in Eq.(8)
one can recognize the complement of the GL kink ({\em i.e.}, the
Davydov $\beta$- kink) playing the role of the probability density
distribution \cite{Can93}. Following Brax, one can develop a multifractal
formalism if the kink probabilistic distribution function is identified with
the $\tau$- function of the multifractal formalism
\begin{equation}\label{eq:r1}
{\cal G}(\xi _{2})- {\cal G}(\xi _{1}) \; \approx
\int _{\xi _{1}}^{\xi _{2}} \beta (\xi ')d\xi '\; \equiv -\tau (\xi)
\end{equation}
I recall that in the multifractal formalism the function $\tau _{q}$
is the cumulant generating function. The derivative of this function
with respect to $q$ is denoted by $\alpha$ and via the Lagrange
multipliers procedure of statistical thermodynamics, one obtains the
function $f(\alpha)$ which is the density of a measure, and could be
interpreted as a fractal dimension when it is positive, and related
to instabilities for negative values. The equations
$\alpha =\delta \tau /\delta q$ and $f(\alpha)=q\alpha -\tau$ represent the
basis
of the multifractal formalism \cite{Man89}. Formally, $q$ is the inverse
temperature, $\tau$ is the Gibbs free energy, and $f$ is the entropy.
One can plot the second derivative of the $\tau$-function (the `specific
heat') and find out intervals of the `temperature' variable, within
which the plot clearly displays features of a phase transition, that
is a peak in the `specific heat' at a certain value of the `temperature'
variable \cite{Can93}.

\section{Concluding Remarks}

The digitally disordered
Davydov beta kinks can display features characteristic to
self-organized criticality.
Here, I considered the digital noise of Davydov kinks as a
reflection of hidden microscopic degrees of freedom of the protein chain.
It is an internal noise in the SOC terminology, representing the
dynamics of a self-organized state of the chain, therefore not destroying the
soliton solution all over the scaling region.

I ascribed the 1/f noise to carrier trapping and so to some sort of
short-distance
thermal activation physics, rather than to coherence effects. This would be
more in the standard spirit of solid state physics. In general 1/f mesoscopic
noise can have many interpretations and it would be helpful to disentangle
its real origin.
At the level of vibrational degrees of freedom it will be
of interest to see if the vibrational soliton possesses self-organized
features as well, and to investigate the connection with the kink digital
disorder.

\section*{Acknowledgments}

The work was supported by the CONACyT Project 4868-E9406.
The author is grateful to Dr. E. Canessa for helpful discussions.

\end{document}